\documentclass[%
unsortedaddress,
preprint,
 amsmath,amssymb,
 aps,
prb,
floatfix,
]{revtex4-2}

\usepackage{graphicx}% Include figure files
\usepackage{dcolumn}% Align table columns on decimal point
\usepackage{bm}% bold math
\usepackage{float}
\begin{document}

\title{Four errors students make with inverse-square law vectors}

\author{Colin S. Wallace}
\affiliation{Department of Physics and Astronomy, University of North Carolina at Chapel Hill}
\email{cswphys@email.unc.edu}

\author{Liam Jones}
\affiliation{Department of Physics and Astronomy, University of North Carolina at Chapel Hill}
\email{ljones15@live.unc.edu}

\author{Alex Lin}
\affiliation{Department of Chemistry, University of Texas at Austin, Austin, Texas 78712}
\email{alexanderlin@utexas.edu}

\date{\today}

\begin{abstract}
In this paper, we discuss four errors introductory physics students make when attempting to add two inverse-square law vectors.  We observe multiple instances in which students 1) add vectors as if they were scalars, 2) project the $r$ (or $r^2$) in the denominator, instead of the entire vector, when attempting to find the vector's components, 3) incorrectly apply the Pythagorean theorem when attempting to calculate the magnitude of the resultant vector, and 4) incorrectly relate the signs of the components of an electric field (or force) to the signs of the electric charges.  While these are not the only errors students make, they are the most frequently occurring based on our analysis of 678 exams taken by students in either introductory mechanics or electricity and magnetism (E\&M).  We then show how these errors can be encoded into a new type of activity or assessment question which we call a ``student error task."  Introductory physics instructors can use the student error task in this paper as a way to engage or assess their students' understandings of how to add two inverse-square law vectors.  

\end{abstract}

\maketitle

\section{Introduction}
\label{intro}
Vectors are the essential mathematical objects of introductory physics.  Students typically experience a number of well-documented difficulties as they learn how to reason with vectors, both in terms of their general mathematical properties \cite{barniol2014test,buncher2015algebra,flores2004student,heckler2015adding,hawkins2009students,hawkins2010students,knight1995vector,mikula2013student,nguyen2003initial,van2007comparing} and their applications to specific physical scenarios \cite{aguirre1984students,aguirre1988student,barniol2013students,liu2021study,rainson1994students,shaffer2005research,southey2014vector,viennot1992students}.  One special type of vector that receives scant attention in the physics education research literature is the inverse-square law.  Three of the most important vectors students encounter in introductory physics take the form of an inverse-square law: Newton's law of gravitation, Coulomb's law, and the electric field of a point charge.  In this paper, we report on four difficulties students experience when working with inverse-square laws, including one that seems to be uniquely triggered by the presence of the $r^2$ in the denominator. We then describe a new type of activity or assessment question, called a ``student error task," which we developed based on our work documenting these four common errors.  Instructors can use and/or adapt this paper's student error task to diagnose their own students' understandings of how to add two inverse-square law vectors.

\section{The Data Set and Question}
\label{data}

\begin{figure}[h]
    \centering
    \includegraphics[height=15cm]{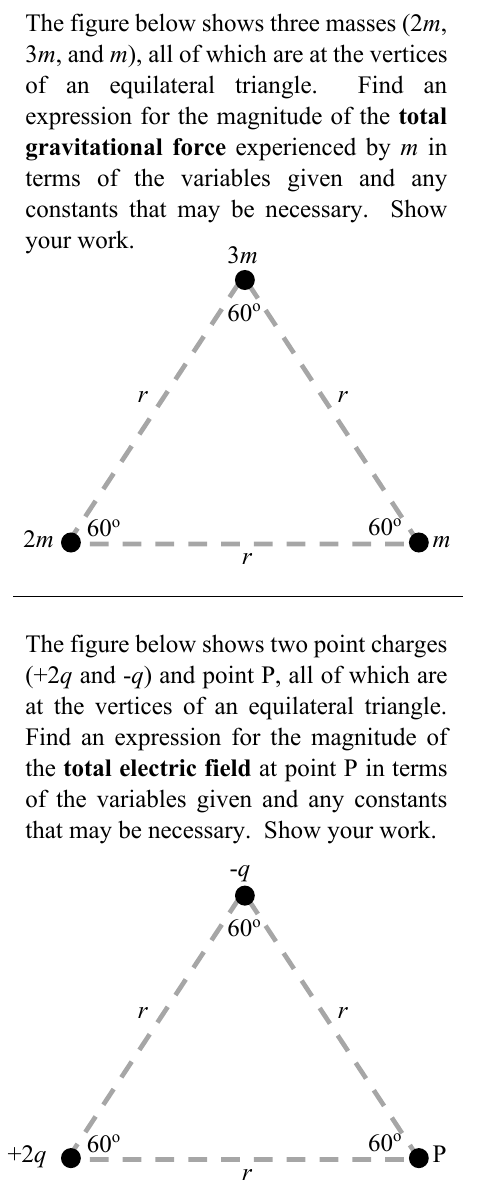}
    \caption{Two examples of the ``equilateral triangle" problem.  The top variant asks for an expression for the total (net) gravitational force experienced by one of the three masses, making this version suitable for a mechanics class studying Newton's law of gravitation.  The bottom variant asks for the total (net) electric field at point P due to two point charges.  This variant is appropriate for a course in electricity and magnetism.}
    \label{fig:FIG1}
\end{figure}

In order to study students' difficulties with inverse-square law vectors, we placed the ``equilateral triangle" problem, variants of which are shown in Figure \ref{fig:FIG1}, on exams taken by students at the University of North Carolina at Chapel Hill (UNC-CH).  These students were enrolled in either the first- or second-semester courses in our introductory sequence for physical science majors (PHYS 118 and 119, respectively) or the second-semester course in our introductory sequence for life science majors (PHYS 115).  In both sequences, mechanics is covered during the first semester and electricity and magnetism are covered in the second.  The first semester of the life science sequence is not represented in our data set due to the fact that Newton's law of gravitation, the only inverse-square law typically encountered in mechanics, is not part of the introductory physics curriculum for life science majors at UNC-CH (see Smith \emph{et al}.\ 2018 for more information on which topics are included and excluded \cite{smith2018transforming}).  Table \ref{tab:demographics} shows the number of exams we analyzed from each course, along with the semester in which the course was taught and the type of exam administered (midterm vs. final exam).

\begin{table}[hbt]
    %\centering
    \begin{ruledtabular}
    \begin{tabular}{lcccl}
    \textrm{Class}&
    \textrm{Content}&
    \textrm{Semester}&
    \textrm{$N$}&
    \textrm{Exam Type}\\
    \colrule
    PHYS 115 & E\&M & Fall 2018 & 137 & midterm \\
    PHYS 115 & E\&M & Fall 2018 & 145 & final \\
    PHYS 118 & Mechanics & Fall 2018 & 166 & final \\
    PHYS 119 & E\&M & Fall 2018 & 44 & midterm \\
    PHYS 118 & Mechanics & Spring 2019 & 91 & final \\
    PHYS 119 & E\&M & Spring 2019 & 95 & midterm\\
    \end{tabular}
    \end{ruledtabular}
    \caption{The classes represented in this data set, their content (either mechanics or E\&M), the semester in which they were taught, the number of student responses ($N$) we collected, and the type of exam on which the equilateral triangle problem appeared.  Note: At UNC-CH, PHYS 115 is the second course in our two-semester sequence for life science majors, while PHYS 118 and 119 are the first and second courses, respectively, in our sequence for physical science majors.}
    \label{tab:demographics}
\end{table}

Readers should note that there are a large number of variants of the equilateral triangle problem that are not shown in Figure \ref{fig:FIG1}.  For example, one could place another charge at point P and ask students to calculate the net electric force felt by one of the charges.  One could also increase or decrease the masses and charges by any arbitrary amount.  One could even change which charges are positive and which are negative.  Any of these changes, or combinations thereof, leads to a new variant of the equilateral triangle problem.  While we restrict our attention in this paper to those variants shown in Figure \ref{fig:FIG1}, researchers and instructors trying to reproduce our findings may wish to use a novel variant.

All the variants of the equilateral triangle question share a solution pathway.  They all ask students to find the magnitude of some vector quantity at one of the vertices.  This vector quantity is the sum of two other vectors, each of which exists due to the presence of some object (either a mass or an electric charge) at the one of the other vertices.  One of these vectors has horizontal and vertical components that are both non-zero, so in order to correctly add the two vectors, students must be able to 
\begin{itemize}
    \item decompose vectors into their $x$- and $y$-components;
    \item assign each component either a ``+" or ``-" sign, depending on the direction in which the vector points; 
    \item sum both $x$-components to find the $x$-component of the total vector, and then do the same for the $y$-components; and
    \item use the total $x$- and total $y$-components in the Pythagorean theorem to find a value or expression for the magnitude of the total vector.
\end{itemize}
Note: We are using ``$x$-direction" and ``$y$-direction" as synonymous with horizontal and vertical, respectively.  Although students are free to define their $x$- and $y$-directions to point in whatever direction they see fit, virtually all students make the $x$-direction horizontal and the $y$-direction vertical.  Consequently, we will stick with this convention for the rest of this paper.

The steps listed above are foundational for the vector analysis techniques taught in introductory physics.  The equilateral triangle problem, in all of its variants, enables instructors to assess whether or not their students have mastered these essential problem-solving skills (albeit in different physical contexts - i.e., gravitational forces or electric forces/fields).

To a physics expert, the questions shown in Figure \ref{fig:FIG1} may seem straightforward. Many instructors expect their students to be able to correctly answer such questions as a result of taking their course -- and approximately 30\% of the exam responses in our data were fully correct.  The remaining $\sim$70\% made one or more errors.  There are numerous locations along the problem-solving pathway where many students take a wrong turn, as we describe in more detail in Sections \ref{error1}-\ref{error4} below. We will elucidate the frequency of a particular error in our data set whenever possible -- but keep in mind that we cannot do so in every case.  Many students have responses that combine multiple errors and/or their written responses do not always provide enough detail for us to confidently infer what they were thinking.  In such cases, we will follow the lead of previous papers, such as Manogue \emph{et al}.\ (2006) \cite{manogue2006ampere}, and be content to simply report that a particular error is observed (and provide a concrete example), without worrying about specifying how common it is.  Physics instructors should be prepared to encounter and address all manners of student difficulties, both common and uncommon, and it is our hope that the errors described below will aid our colleagues in this endeavor.

\section{Error 1: Treating Vectors as Scalars}
\label{error1}

\begin{figure*}[ht]
    \fbox{\includegraphics[height=8cm]{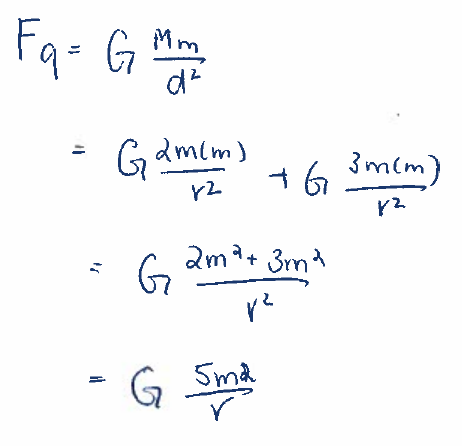}}
  \hfill
    \fbox{\includegraphics[height=5cm]{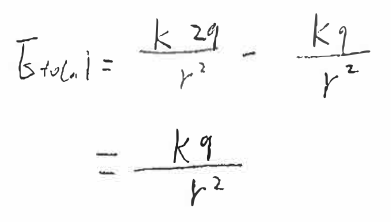}}
  \caption{Two examples of responses that add vectors as scalars.  Top box: An example from mechanics.  The student attempted to find the net gravitational force by adding the magnitudes of two gravitational forces.  Bottom box: An example from E\&M.  The student attempted to find the net electric field by adding the electric fields produced by a $+2q$ and a $-q$ charge.  Notice that the student placed a negative sign in front of the electric field due to the $-q$ charge.  This approach is discussed further in Section \ref{error4}.}
  \label{fig:FIG2}
\end{figure*}

A disheartening number of students do not even attempt to use any vector mathematics.  Close to 18\% of exams in our data set contain responses in which the student adds two vector quantities as if they were scalars.  Figure \ref{fig:FIG2} shows two examples, one from mechanics and one from electricity and magnetism.  This error seems to be more prevalent in mechanics, appearing in 32\% of responses for both semesters of PHYS 118 represented in our data set.  It appears to be less prevalent in the sequel course, PHYS 119, where only 9\% of students attempt to add the vectors as scalars. 
%(although we must avoid the temptation to attribute too much significance to this exact value, given the relatively small number of students in PHYS 119)
In PHYS 115, the electricity and magnetism course for life science majors, 7\% add them as scalars on a midterm exam and 10\% do the same on a similar question on the final exam.  Every student who made this error on the midterm also made the same error on the final exam, despite the fact that the midterm exams were returned to students after they were graded and full midterm solutions were posted for the class to review.  The fact that some PHYS 119 students continue to add vectors as scalars, despite working extensively with vector quantities in PHYS 118, and that some PHYS 115 students make the same mistake on both a midterm and final exam, suggests that there is a non-negligible population of students that never learn to recognize forces and fields as vectors and/or fail to learn how to add vectors.   

\section{Error 2: Projecting $r$ and Not the Entire Vector}
\label{error2}

\begin{figure}[ht]
    \fbox{\includegraphics[height=7cm]{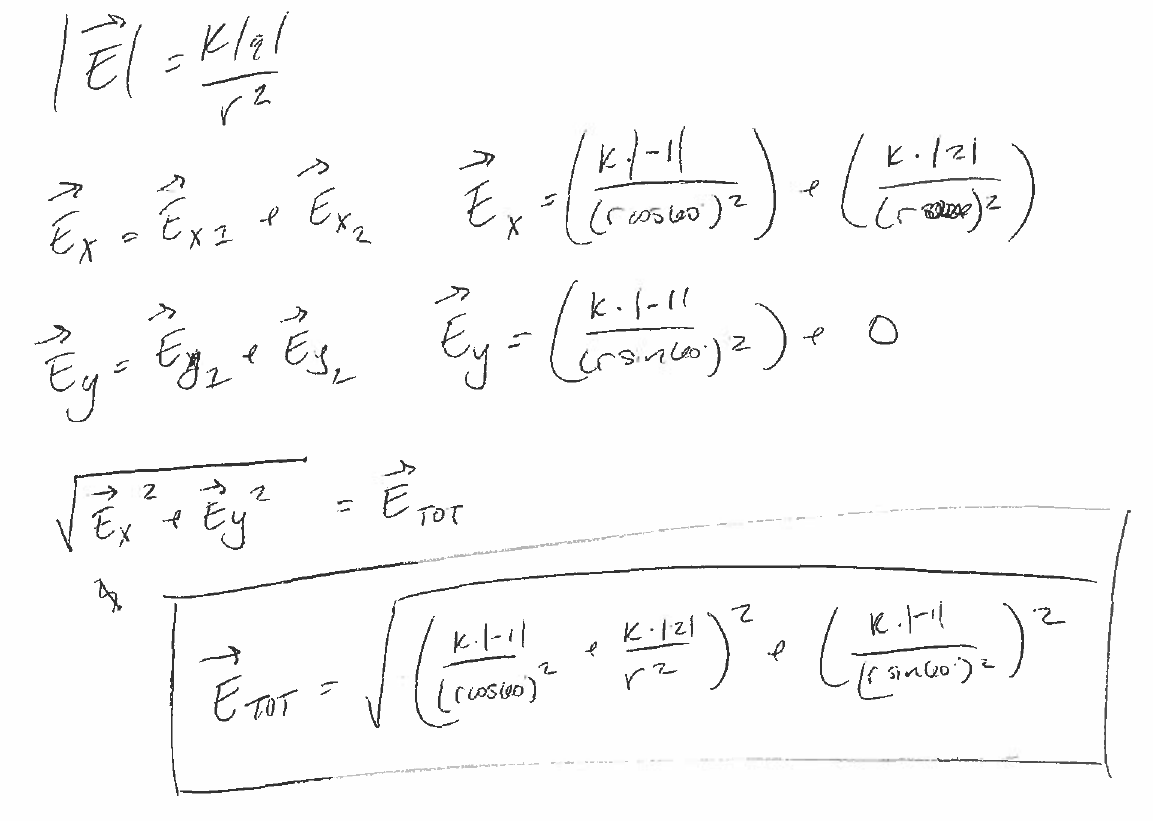}}
  \hfill
    \fbox{\includegraphics[height=6cm]{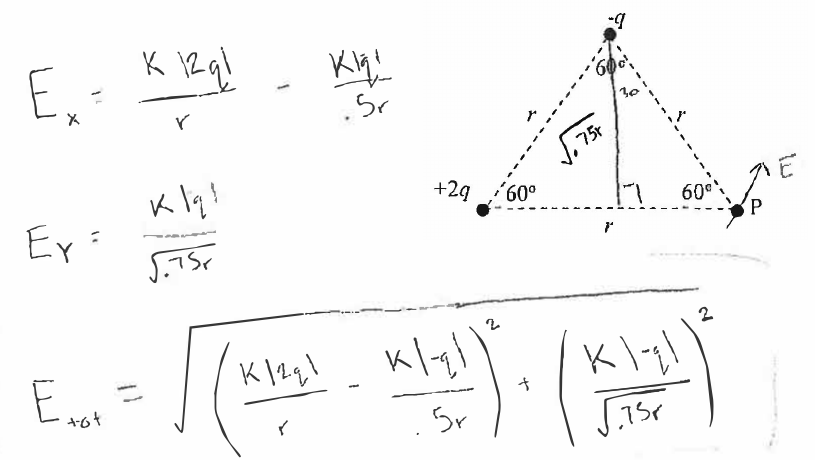}}
  \caption{Two examples of responses that project $r$ instead of the overall vector.  Top box: The student who wrote this response attempted to find the $x-$ and $y-$components of an electric field by multiplying the $r$ in the denominator by cos($60^o$) and sin($60^o$), respectively, resulting in these trigonometric terms appearing in the denominator.  Bottom box: The student who wrote this response drew a vertical line from the triangle's top vertex (where the $-q$ charge is located) down to the bottom of the triangle, bifurcating the bottom into two segments, each of length $r/2$.  They concluded that the $x-$component of $-q$'s electric field has $0.5r$ in the denominator (note that they forgot to square $r$).  This student also seems to have used the Pythagorean theorem to try to determine that the vertical displacement between $-q$ and point P.  They write this displacement as $\sqrt{0.75r}$ and they place this term in the denominator of the $y-$component of $-q$'s electric field.}
  \label{fig:FIG3}
\end{figure}

Experts see quantities such as $GM_1 M_2/r^2$ and $kq/r^2$ as the magnitudes of vectors (in these cases, a gravitational force and the electric field of a point charge, respectively).  When asked to find a component of one of these vectors, experts will multiply the magnitude by either the sine or cosine of some angle.  In contrast, some students appear to think that the $r$ (or $r^2$) in the denominator is, by itself, the vector.  By this reasoning, the expressions for the gravitational force and the electric field of a point charge are only vectors because the $r^2$ is present.  Consequently, when these students attempt to find a component of an inverse-square law vector, they will only multiply $r$ or $r^2$ by the sine or cosine term, resulting in an answer that has these trigonometric quantities in the denominator.  Figure \ref{fig:FIG3} shows an example.  We can succinctly describe this particular error by saying that students are projecting just $r$ (or $r^2$) on a certain dimension and not the entire vector.  Since inverse-square laws are the most prominent example of vectors with an $r$ in the denominator, they are well-suited for evoking this particular misunderstanding.

Figure \ref{fig:FIG3} also shows a variant of this error.  Some students do not explicitly place a trigonometric term in the denominator.  Instead, they draw a vertical line that starts at the top vertex and extends to the bottom of the triangle, bifurcating the bottom line into two equal segments of length $r/2$.  They then claim that the $x$-component of a given vector will have $0.5r$ in the denominator. As an example, the bottom panel of Figure \ref{fig:FIG3} shows the work of a student trying to figure out the net electric field at point P shown in Figure \ref{fig:FIG1}.  This student recognized that they needed to find the $x-$ and $y-$components of the electric field due to the charge $-q$.  Since $-q$ has a horizontal displacement of $r/2$ from point P, the student claimed that the $x-$component of $-q$'s electric field has a magnitude of $kq/(0.5r)$  Note that this student apparently forgot to square the distance term that appears in the denominator.

The bottom panel of Figure \ref{fig:FIG3} also shows that this student concluded that the $y-$component of $-q$'s electric field is given by $kq/\sqrt{0.75r}$.  They likely arrived at the $\sqrt{0.75r}$ via the Pythagorean theorem.  If the hypotenuse of a triangle is $r$ and one leg has a length $0.5r$, then one can reconstruct from the Pythagorean theorem how the student arrived at the conclusion that the other leg, represented by their vertical line, has a length $\sqrt{0.75r}$.  Note that this specific result also requires an additional error, namely that one also mistakenly use $0.75r$ in place of $0.75r^2$.  Regardless of any confusion between $0.75r$ and $0.75r^2$, this student attempted to determine the $y-$component of $-q$'s electric field by inserting into the denominator the vertical displacement between $-q$ and point P.  Responses from other students also use the this kind of approach to determine the $y-$component.

The two approaches shown in Figure \ref{fig:FIG3} may lead to mathematically equivalent results (e.g., dividing by cos($60^o$) is the same as dividing by 0.5).  That does not mean that students necessarily see these approaches as equivalent.  We suspect that many do not, as evidenced by the fact that many students who draw the vertical line ultimately do not multiply $r$ by a sine or cosine, preferring instead to express the denominator as some fraction of $r$.  We suspect that students who place a sine or cosine in the denominator remember that many physics problems use trigonometric terms to find vector components.  In contrast, students who draw a vertical line may be recalling that the $x-$ and $y-$components of a vector are often drawn as the legs of a right triangle.  They thus incorrectly conclude that the $x-$component and $y-$component of the vector only depend on the $x-$component and $y-$component, respectively, of the displacement between between the two points of interest.  While we believe these are reasonable hypotheses to explain the observed pattern of responses, our data set is not sufficient to confirm or reject these ideas.  Additionally, there are some responses (such as the one discussed in Section \ref{combo}) that seem to mix these two approaches, suggesting that some students have more complex ideas about how break inverse-square laws into components.

How common are these errors?  For any given exam, we typically see that approximately 5-15\% of responses project $r$ (or $r^2$) rather than the entire vector, either by placing a trigonometric term in the denominator or by using the $x-$ or $y-$displacement in place of $r$ (or, in a small number of cases, some combination of both).  These errors are most common in the PHYS 115 exams, where they occur in 16\% of the responses we examined.  They are least common in PHYS 119; only 6\% of these responses exhibit these errors.  PHYS 118 is in the middle at 9\%.  These results seem to make sense: Students who struggle with vectors in PHYS 118 (or with the course content in general) are less likely to take PHYS 119.  PHYS 119 also has vector calculus as a co-requisite, so we should not be surprised if these students are more comfortable with vector mathematics (but see Section \ref{error1} for information suggesting that approximately 9\% of these students may try to add the vectors as scalars). Furthermore, PHYS 115, unlike PHYS 118 and 119, has no calculus prerequisite.  We therefore expect a higher percentage of PHYS 115 students will struggle with these mathematical representations.

\section{Error 3: Incorrectly Applying the Pythagorean Theorem}
\label{error3}

\begin{figure}[ht]
    \fbox{\includegraphics[height=8cm]{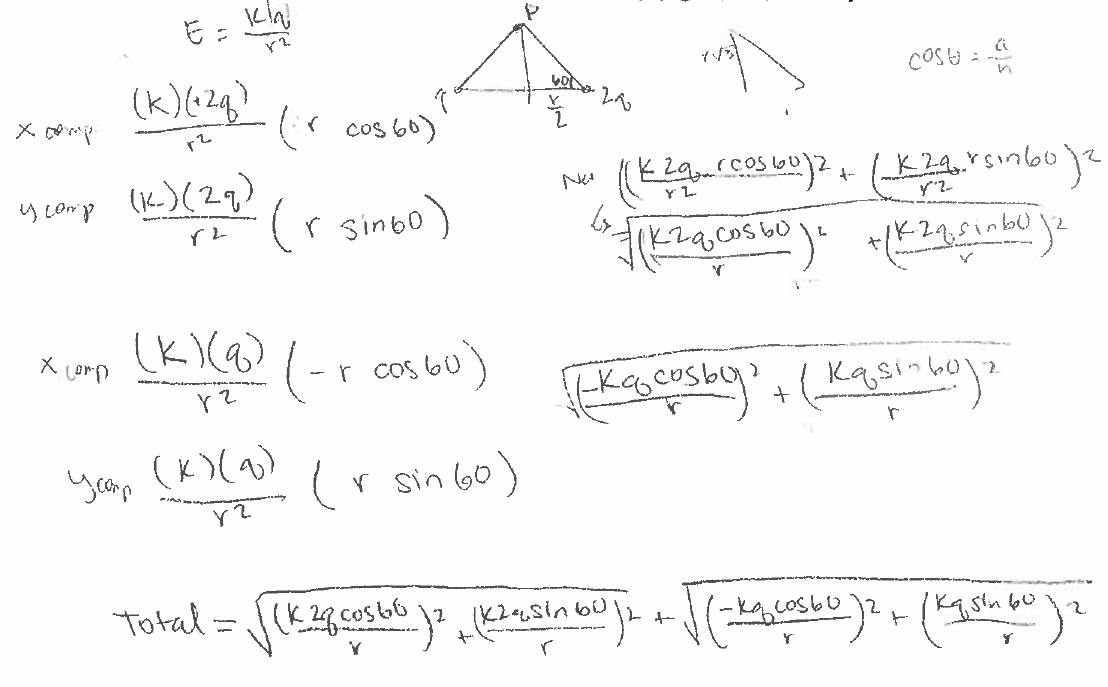}}
  \hfill
    \fbox{\includegraphics[height=7cm]{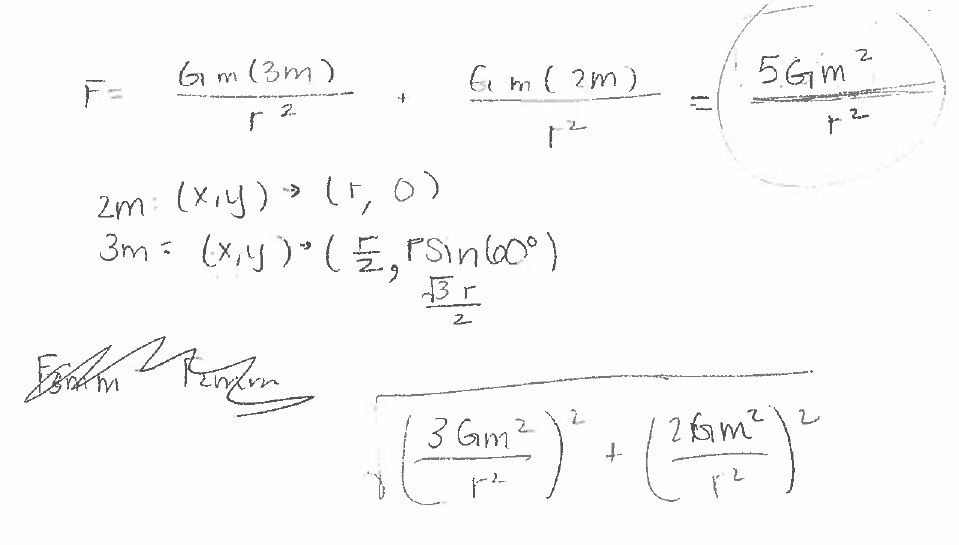}}
  \caption{Two examples of responses that incorrectly apply the Pythagorean theorem.  Top box: The student who wrote this response attempted to find the $x-$ and $y-$components of the net electric field by adding in quadrature the $x-$ and $y-$components, respectively, of the individual electric fields.  They then claimed that the net electric field is the sum of the net field's $x-$ and $y-$components.  Bottom box: The student who wrote this response attempted to find the net gravitational force by adding the magnitudes of two gravitational forces in quadrature.  Note that this student also tried to add the two forces as if they were scalars, an error discussed in Section \ref{error1}.}
  \label{fig:FIG4}
\end{figure}

While many students know that they must use the Pythagorean theorem to find the magnitude of the net force/field vector, they do not all invoke the Pythagorean theorem at the appropriate point in the problem-solving process.  Figure \ref{fig:FIG4} shows two examples.  The example in the top box shows the work of a student who attempted to calculate the total $x-$component by placing the $x-$components of the individual vectors in the Pythagorean theorem (and likewise for the $y-$components).  The bottom box is an example of a response in which the student attempted to calculate the magnitude of the net gravitational force by inserting into the Pythagorean theorem the magnitudes of the two individual forces (note that this student also attempted to solve the problem by adding the two forces as scalars).  Responses such as these suggest that some students invoke Pythagoras as part of a partially-remembered algorithm, but lack a complete understanding of how and why to apply it.  These students are activating what Reif (1987) would call inappropriate or mis-remembered procedural knowledge \cite{reif1987interpretation}.

\section{Error 4: Incorrectly Relating the Signs of Components to the Signs of Electric Charges}
\label{error4}

\begin{figure}[ht]
    \fbox{\includegraphics[height=10cm]{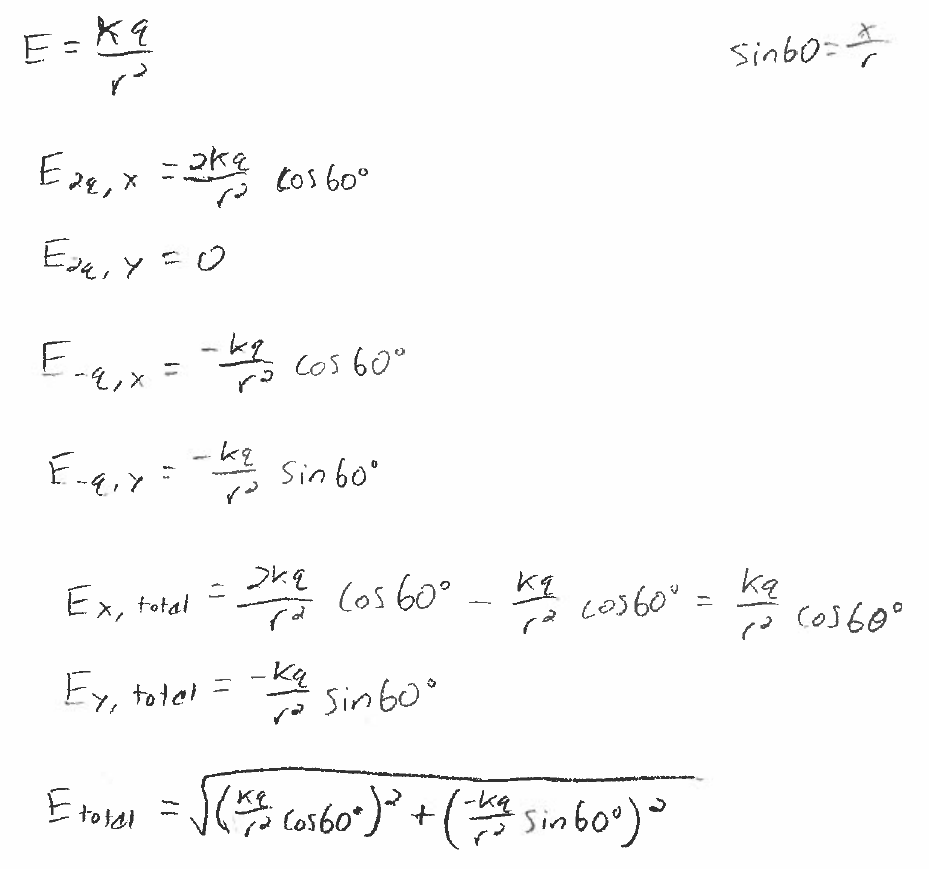}}
  \hfill
    \fbox{\includegraphics[height=7cm]{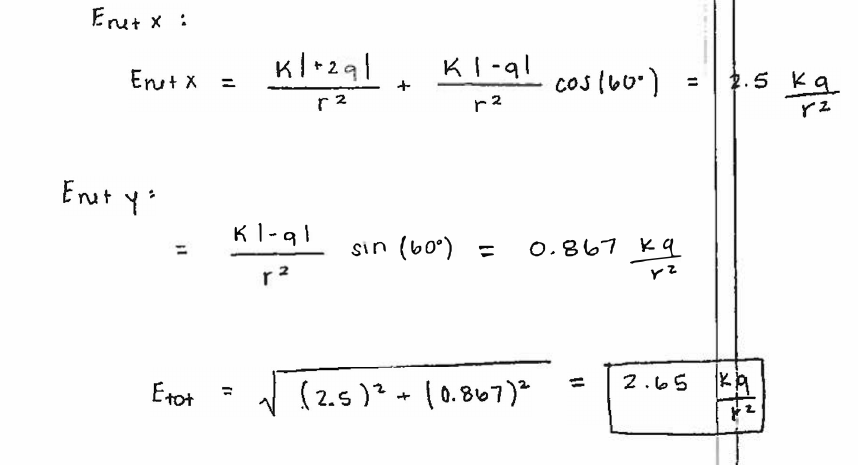}}
  \caption{Two examples of how students attempt to deal with negative charges.  Top box: The student who wrote this response placed a negative sign in front of both components of the electric field of the charge $-q$.  Bottom box: The student who wrote this response took the absolute value of all charges (positive and negative), rendering all terms positive.}
  \label{fig:FIG5}
\end{figure}

This final error only occurs in problems about electric fields or forces.  While the sign of a charge does determine the direction of its electric field and the electric force it exerts on and feels from another charge, students do not always know how to encode this knowledge in a mathematical expression.  An expert might use their conceptual knowledge to determine the direction of the field or force vector and then assign signs to this vector's $x-$ and $y-$components based on how they have defined the positive and negative $x$ and $y$ directions for their coordinate system.  In contrast, many students employ a less sophisticated approach.  Some will automatically include a negative sign with any component of any vector associated with a negative charge, regardless of whether or not that vector points in a direction they (implicitly or explicitly) treat as positive elsewhere in their solution.  An example of this is shown in the top box of Figure \ref{fig:FIG5}.  Others will correctly recall that the magnitude of an electric force or field only depends on the absolute value(s) of the relevant charge(s) -- but they will then place absolute value signs around all charges that appear in these expressions, rendering all quantities positive.  The bottom box of Figure \ref{fig:FIG5} shows an example of this approach.

This can be a difficult error to detect, especially in students' written solutions to exam questions.  There are many situations in which these incorrect procedures can lead to results that are mathematically correct.  A component of a negative charge's electric field can be either positive or negative, depending on the orientation of the field and how one has defined the positive and negative directions on one's coordinate system.  When examining our data set, we found many examples where we could not be sure whether students had actually made an error, which is why we do not attempt to quantify the prevalence of either error described in this section.  Researchers interested in such quantification should consider supplementing written responses with ``think aloud" interviews that may provide additional insights into students' reasoning \cite{otero2009getting,willis2005cognitive}.

\section{Combinations of Errors}
\label{combo}

\begin{figure}[ht]
    \fbox{\includegraphics[height=5cm]{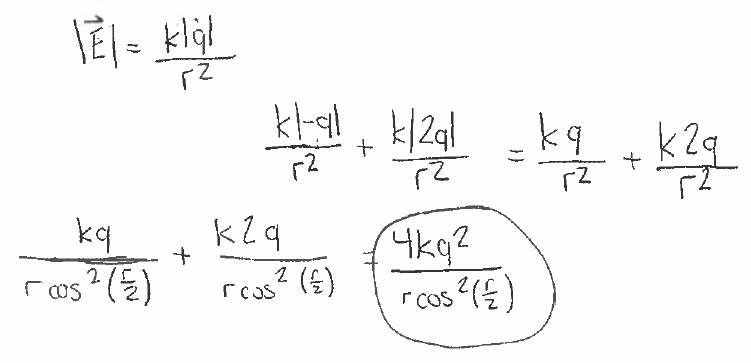}}
  \caption{A sample response that exhibits multiple errors.  The student who wrote this response took the absolute value of both charges, placed a trigonometric term (cosine) in the denominator, and made $r/2$ the argument of the cosine.  The fact that the response contains both a cosine and $r/2$ suggests that the student was only reasoning about the $x-$components of these vectors.}
  \label{fig:FIG6}
\end{figure}

There are plenty of responses that exhibit more than one of the errors described in Sections \ref{error1}-\ref{error4}.  We already saw an example of this in the bottom box of Figure \ref{fig:FIG4}, which shows the response of a student who tried to find the net gravitational force by both 1) adding the individual force vectors as scalars and 2) inserting the individual force vectors into the Pythagorean theorem.  Figure \ref{fig:FIG6} shows another example.  In this response, the student 1) placed the charges in absolute values, making all quantities positive; 2) tried adding the vectors as scalars; 3) placed a cosine term in both denominators (suggesting that they might have decided to focus solely on $x-$components, ignoring the $y-$components completely); and 4) made $r/2$ the argument of their cosine terms.  While this particular combination of errors is idiosyncratic to this student, their response highlights the fact that instructors must be prepared to identify multiple errors in a single response.

\section{Student Error Tasks -- An Alternative Type of Exam Questions}
\label{exam}

\begin{figure}[h]
    \centering
    \includegraphics[height=8cm]{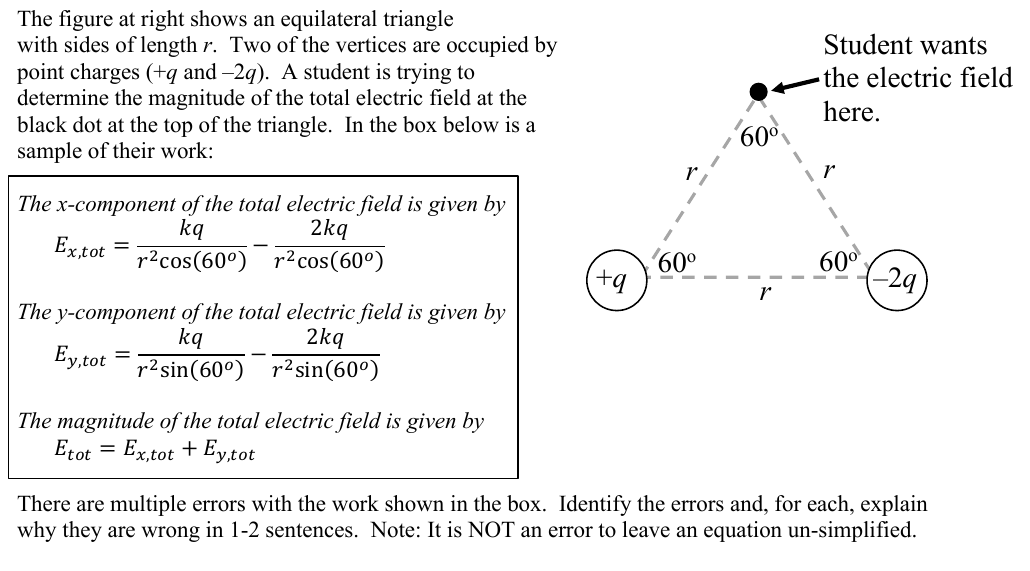}
    \caption{An example student error task.  This question asks students to identify the errors made by a fictional student.  The errors made by this fictional student correspond to some of the errors we identified in the responses of real students.}
    \label{fig:FIG7}
\end{figure}

Our work in identifying the errors described in Sections \ref{error1}-\ref{error4} inspired us to develop the ``student error task," a type of exam question that we had never before used.  A student error task presents students with a snippet of a fictional student's work on a quantitative problem.  This snippet contains one or more errors, which our real students must identify.  Figure \ref{fig:FIG7} shows the student error task we created as a result of the information we learned from this study.  Although we developed the idea for student error tasks on our own, a subsequent review of the literature shows that others previously created similar tasks \cite{barbieri2020mistakes,de2023self,maclean2023s,yerushalmi2006guiding}.

In the example shown in Figure \ref{fig:FIG7}, there are three errors students must identify in order to receive full credit:
\begin{enumerate}
    \item In the term for $E_{x,tot}$, both terms should have the same sign.  (Most students -- perhaps most people -- default to a coordinate system in which the $+x$-direction points to the right.  In this case, the minus sign in front of the second term should be positive.  However, it is equally valid to define the $-x$-direction as the one that points right, in which case both terms in the expression for $E_{x,tot}$ should be negative.  Few students will define $-x$ as pointing to the right, but any that do will receive full credit if they also argue that both terms should be negative.)
    \item The $\cos(60^o)$ and $\sin(60^o)$ terms should be in the numerators, not denominators, of their respective terms.
    \item The magnitude of the total electric field is given by the Pythagorean theorem, $E_{tot} = \sqrt{E_{x,tot}^2 + E_{y,tot}^2}$, not the sum of the components.
\end{enumerate}
These errors all derive from errors we observed real students make, as described in the preceding sections.

Note that we do not show a full solution in the snippet.  For example, one may wonder why we write the total $x$-component as
\begin{equation*}
E_{x,tot} = \frac{kq}{r^2 \cos(60^o)}-\frac{2kq}{r^2 \cos(60^o)}
\end{equation*}
instead of simplifying this (incorrect) expression to
\begin{equation*}
E_{x,tot} = -\frac{kq}{r^2 \cos(60^o)}.
\end{equation*}
This is intentional.  If we wrote $E_{x,tot} = -\frac{kq}{r^2 \cos(60^o)}$ then, rather than spending their time identifying errors, students might spend time trying to reproduce this incorrect expression.  If they fail to reproduce this expression then we learn nothing about their ability to identify and explain the errors contained within.  By keeping $E_{x,tot}$ as $\frac{kq}{r^2 \cos(60^o)}-\frac{2kq}{r^2 \cos(60^o)}$, we enable students to more readily recognize that each term is the purported $x$-component of one of the relevant vectors, which facilitates identifying the error in the relative signs of these terms.  

The first time we put the student error task in Figure \ref{fig:FIG7} on an exam was during the spring 2023 semester.  This question appeared on a midterm exam for students taking PHYS 115.  358 students took the exam.  The vast majority of students did not try to solve the problem for themselves, instead choosing to use their time to critique to provided solution, as directed in the prompt.  This was the first time students encountered a student error task.  The mean score on this question was 60\% (median: 70\%), with a standard deviation of 26\%.  13\% of students correctly identified all three errors and earned full credit.  Of those that did not receive full credit,  75\% recognized that the calculation of $E_tot$ from its components is incorrect, 35\% recognized that the trigonometric terms should not be in the denominator, and 20\% recognized that the two $x$-components should be added together, not subtracted. 

Although the PHYS 115 students had not previously encountered a student error tasks \emph{per se}, they had encountered the question type known as ``student debates," which are used in many active learning curricula \cite{mcdermott2002tutorials,prather2022lecture}, including the ``Physics Activities for the Life Sciences," which were developed for UNC-CH's life science sequence \cite{smith2018transforming}.  A traditional student debate activity shows the responses of two or more fictional students to a question.  The real students doing the activity must then explain which of the fictional students, if any, is correct and why.  While student error tasks only contain the work of a single fictional student, the cognitive engagement of a student error task is similar to that of a student debate.  Both require students to operate at the ``evaluation" level of Bloom's taxonomy \cite{bloom2020taxonomy}.

Even though we have (to date) only used student error tasks on exams, their similarities to student debates suggest that they could also be used as in-class, collaborative activities.  Identifying errors is a slightly different, but complementary, challenge compared to traditional problem-solving exercises.  Both may be important.  Together, they may help students learn which problem-solving steps are correct and which are not.  More importantly, they make students \emph{think} about what does and does not work.  This suggests that student error tasks, used in tandem with traditional problem-solving exercises (which can be done interactively, even in a large-enrollment class \cite{wallace2021students,wallace2020developing}), may foster the development of students' metacognitive skills.  Metacognition is essential for successful problem solving and prior studies show that it is one of the key differences between novices and experts \cite{schoenfeld1987s}.

Other physics instructors may be interested in developing and implementing their own student error tasks, either for exams or as in-class activities (or both).  We present the example shown in this paper to demonstrate how instructors can translate observations of students' difficulties into practical instructional and assessment activities.

\section{Conclusions}
\label{conclusions}

Inverse-square laws are a small subset of the vector quantities that introductory physics students encounter.  Yet they are important.  Some of the most fundamental relationships in physics are inverse-square laws.  Additionally, the act of vectorially adding two-inverse square law vectors requires students to confront many foundational principles of vector addition, including how to break vectors into components, when to apply the Pythagorean theorem, and (in situations involving electric charges) how to relate the signs of charges to the signs of the components of electric fields and forces.  We have documented student difficulties with all of these principles.

The common student errors we identified are important for instructors to keep in mind, even when they are teaching topics other than inverse-square laws.  Many of these errors suggest fundamental misunderstandings of how to translate a physical scenario into appropriate mathematical relationships.  For example, if we connected the masses in Figure \ref{fig:FIG1} by springs and asked students to add two spring forces, we expect that many students will continue to add these vectors as scalars and/or incorrectly apply the Pythagorean theorem.

The ability to translate physics into math is a critical one for students to develop if they are to become successful problem solvers, a goal of many physics classes.  The literature on problem solving is vast \cite{hsu2004resource}.  Prior research shows that one of, if not the most, significant stumbling block for students is their inability to activate the appropriate conceptual knowledge and encode it into mathematical expressions, instead resorting (often unsuccessfully) to incompletely-remembered procedural steps and algorithms \cite{reif1987interpretation,leonard1996using,sabella2007knowledge,wilcox2013analytic}.  Any instructor who has heard their students say ``I just don't know where to begin" will recognize this issue.  The student error task we provide in this paper is one tool instructors can use, either as an in-class activity to help students recognize and overcome common errors or as an assessment question to diagnose the prevalence of these errors.  Instructors may also be inspired to develop their own student error tasks based on other student difficulties documented in the physics education research literature and/or based on observations of their own students.  We hope the information we present in this paper will help improve the teaching and learning of these fundamentally important ideas.

\section{Acknowledgements}
\label{acknowledge}
We thank our colleagues who, over the years, have discussed this work with us and who let us place the equilateral triangle problem on their exams.  We especially appreciate the assistance of Alice Churukian, Duane Deardorff, Stefan Jeglinski, Jennifer Weinberg-Wolf, and Dan Young.

\section{Data Availability}
\label{data}
The data that support the findings of this study are available upon request from the authors.

\section{Author Declarations}
\label{declarations}
The authors have no conflicts to disclose.  This work was approved by the University of North Carolina at Chapel Hill's Institutional Review Board (IRB No.\ 18-1276).  This research was conducted in accordance with the principles embodied in the Declaration of Helsinki and in accordance with local statutory requirements.  All participants gave written informed consent to participate in the study.

\bibliography{bibliography}% Produces the bibliography via BibTeX.

\begin{thebibliography}{37}%
\makeatletter
\providecommand \@ifxundefined [1]{%
 \@ifx{#1\undefined}
}%
\providecommand \@ifnum [1]{%
 \ifnum #1\expandafter \@firstoftwo
 \else \expandafter \@secondoftwo
 \fi
}%
\providecommand \@ifx [1]{%
 \ifx #1\expandafter \@firstoftwo
 \else \expandafter \@secondoftwo
 \fi
}%
\providecommand \natexlab [1]{#1}%
\providecommand \enquote  [1]{``#1''}%
\providecommand \bibnamefont  [1]{#1}%
\providecommand \bibfnamefont [1]{#1}%
\providecommand \citenamefont [1]{#1}%
\providecommand \href@noop [0]{\@secondoftwo}%
\providecommand \href [0]{\begingroup \@sanitize@url \@href}%
\providecommand \@href[1]{\@@startlink{#1}\@@href}%
\providecommand \@@href[1]{\endgroup#1\@@endlink}%
\providecommand \@sanitize@url [0]{\catcode `\\12\catcode `\$12\catcode
  `\&12\catcode `\#12\catcode `\^12\catcode `\_12\catcode `\%12\relax}%
\providecommand \@@startlink[1]{}%
\providecommand \@@endlink[0]{}%
\providecommand \url  [0]{\begingroup\@sanitize@url \@url }%
\providecommand \@url [1]{\endgroup\@href {#1}{\urlprefix }}%
\providecommand \urlprefix  [0]{URL }%
\providecommand \Eprint [0]{\href }%
\providecommand \doibase [0]{https://doi.org/}%
\providecommand \selectlanguage [0]{\@gobble}%
\providecommand \bibinfo  [0]{\@secondoftwo}%
\providecommand \bibfield  [0]{\@secondoftwo}%
\providecommand \translation [1]{[#1]}%
\providecommand \BibitemOpen [0]{}%
\providecommand \bibitemStop [0]{}%
\providecommand \bibitemNoStop [0]{.\EOS\space}%
\providecommand \EOS [0]{\spacefactor3000\relax}%
\providecommand \BibitemShut  [1]{\csname bibitem#1\endcsname}%
\let\auto@bib@innerbib\@empty
%</preamble>
\bibitem [{\citenamefont {Barniol}\ and\ \citenamefont
  {Zavala}(2014)}]{barniol2014test}%
  \BibitemOpen
  \bibfield  {author} {\bibinfo {author} {\bibfnamefont {P.}~\bibnamefont
  {Barniol}}\ and\ \bibinfo {author} {\bibfnamefont {G.}~\bibnamefont
  {Zavala}},\ }\bibfield  {title} {\bibinfo {title} {Test of understanding of
  vectors: A reliable multiple-choice vector concept test},\ }\href@noop {}
  {\bibfield  {journal} {\bibinfo  {journal} {Physical Review Special
  Topics-Physics Education Research}\ }\textbf {\bibinfo {volume} {10}},\
  \bibinfo {pages} {010121} (\bibinfo {year} {2014})}\BibitemShut {NoStop}%
\bibitem [{\citenamefont {Buncher}(2015)}]{buncher2015algebra}%
  \BibitemOpen
  \bibfield  {author} {\bibinfo {author} {\bibfnamefont {J.~B.}\ \bibnamefont
  {Buncher}},\ }\bibfield  {title} {\bibinfo {title} {Algebra-based students
  and vector representations: Arrow vs. ijk},\ }\href@noop {} {\bibfield
  {journal} {\bibinfo  {journal} {American Association of Physics Teachers}\ ,\
  \bibinfo {pages} {75}} (\bibinfo {year} {2015})}\BibitemShut {NoStop}%
\bibitem [{\citenamefont {Flores}\ \emph {et~al.}(2004)\citenamefont {Flores},
  \citenamefont {Kanim},\ and\ \citenamefont {Kautz}}]{flores2004student}%
  \BibitemOpen
  \bibfield  {author} {\bibinfo {author} {\bibfnamefont {S.}~\bibnamefont
  {Flores}}, \bibinfo {author} {\bibfnamefont {S.~E.}\ \bibnamefont {Kanim}},\
  and\ \bibinfo {author} {\bibfnamefont {C.~H.}\ \bibnamefont {Kautz}},\
  }\bibfield  {title} {\bibinfo {title} {Student use of vectors in introductory
  mechanics},\ }\href@noop {} {\bibfield  {journal} {\bibinfo  {journal}
  {American Journal of Physics}\ }\textbf {\bibinfo {volume} {72}},\ \bibinfo
  {pages} {460} (\bibinfo {year} {2004})}\BibitemShut {NoStop}%
\bibitem [{\citenamefont {Heckler}\ and\ \citenamefont
  {Scaife}(2015)}]{heckler2015adding}%
  \BibitemOpen
  \bibfield  {author} {\bibinfo {author} {\bibfnamefont {A.~F.}\ \bibnamefont
  {Heckler}}\ and\ \bibinfo {author} {\bibfnamefont {T.~M.}\ \bibnamefont
  {Scaife}},\ }\bibfield  {title} {\bibinfo {title} {Adding and subtracting
  vectors: The problem with the arrow representation},\ }\href@noop {}
  {\bibfield  {journal} {\bibinfo  {journal} {Physical Review Special
  Topics-Physics Education Research}\ }\textbf {\bibinfo {volume} {11}},\
  \bibinfo {pages} {010101} (\bibinfo {year} {2015})}\BibitemShut {NoStop}%
\bibitem [{\citenamefont {Hawkins}\ \emph {et~al.}(2009)\citenamefont
  {Hawkins}, \citenamefont {Thompson},\ and\ \citenamefont
  {Wittmann}}]{hawkins2009students}%
  \BibitemOpen
  \bibfield  {author} {\bibinfo {author} {\bibfnamefont {J.~M.}\ \bibnamefont
  {Hawkins}}, \bibinfo {author} {\bibfnamefont {J.~R.}\ \bibnamefont
  {Thompson}},\ and\ \bibinfo {author} {\bibfnamefont {M.~C.}\ \bibnamefont
  {Wittmann}},\ }\bibfield  {title} {\bibinfo {title} {Students consistency of
  graphical vector addition method on 2-d vector addition tasks},\ }in\
  \href@noop {} {\emph {\bibinfo {booktitle} {AIP Conference Proceedings}}},\
  Vol.\ \bibinfo {volume} {1179}\ (\bibinfo {organization} {American Institute
  of Physics},\ \bibinfo {year} {2009})\ pp.\ \bibinfo {pages}
  {161--164}\BibitemShut {NoStop}%
\bibitem [{\citenamefont {Hawkins}\ \emph {et~al.}(2010)\citenamefont
  {Hawkins}, \citenamefont {Thompson}, \citenamefont {Wittmann}, \citenamefont
  {Sayre},\ and\ \citenamefont {Frank}}]{hawkins2010students}%
  \BibitemOpen
  \bibfield  {author} {\bibinfo {author} {\bibfnamefont {J.~M.}\ \bibnamefont
  {Hawkins}}, \bibinfo {author} {\bibfnamefont {J.~R.}\ \bibnamefont
  {Thompson}}, \bibinfo {author} {\bibfnamefont {M.~C.}\ \bibnamefont
  {Wittmann}}, \bibinfo {author} {\bibfnamefont {E.~C.}\ \bibnamefont
  {Sayre}},\ and\ \bibinfo {author} {\bibfnamefont {B.~W.}\ \bibnamefont
  {Frank}},\ }\bibfield  {title} {\bibinfo {title} {Students’ responses to
  different representations of a vector addition question},\ }in\ \href@noop {}
  {\emph {\bibinfo {booktitle} {AIP Conference Proceedings}}},\ Vol.\ \bibinfo
  {volume} {1289}\ (\bibinfo {organization} {American Institute of Physics},\
  \bibinfo {year} {2010})\ pp.\ \bibinfo {pages} {165--168}\BibitemShut
  {NoStop}%
\bibitem [{\citenamefont {Knight}(1995)}]{knight1995vector}%
  \BibitemOpen
  \bibfield  {author} {\bibinfo {author} {\bibfnamefont {R.~D.}\ \bibnamefont
  {Knight}},\ }\bibfield  {title} {\bibinfo {title} {The vector knowledge of
  beginning physics students},\ }\href@noop {} {\bibfield  {journal} {\bibinfo
  {journal} {The Physics Teacher}\ }\textbf {\bibinfo {volume} {33}},\ \bibinfo
  {pages} {74} (\bibinfo {year} {1995})}\BibitemShut {NoStop}%
\bibitem [{\citenamefont {Mikula}\ and\ \citenamefont
  {Heckler}(2013)}]{mikula2013student}%
  \BibitemOpen
  \bibfield  {author} {\bibinfo {author} {\bibfnamefont {B.~D.}\ \bibnamefont
  {Mikula}}\ and\ \bibinfo {author} {\bibfnamefont {A.~F.}\ \bibnamefont
  {Heckler}},\ }\bibfield  {title} {\bibinfo {title} {Student difficulties with
  trigonometric vector components persist in multiple student populations},\
  }in\ \href@noop {} {\emph {\bibinfo {booktitle} {Proceedings of the 2013
  Physics Education Research Conference, Portland, OR}}}\ (\bibinfo {year}
  {2013})\ pp.\ \bibinfo {pages} {253--256}\BibitemShut {NoStop}%
\bibitem [{\citenamefont {Nguyen}\ and\ \citenamefont
  {Meltzer}(2003)}]{nguyen2003initial}%
  \BibitemOpen
  \bibfield  {author} {\bibinfo {author} {\bibfnamefont {N.-L.}\ \bibnamefont
  {Nguyen}}\ and\ \bibinfo {author} {\bibfnamefont {D.~E.}\ \bibnamefont
  {Meltzer}},\ }\bibfield  {title} {\bibinfo {title} {Initial understanding of
  vector concepts among students in introductory physics courses},\ }\href@noop
  {} {\bibfield  {journal} {\bibinfo  {journal} {American Journal of Physics}\
  }\textbf {\bibinfo {volume} {71}},\ \bibinfo {pages} {630} (\bibinfo {year}
  {2003})}\BibitemShut {NoStop}%
\bibitem [{\citenamefont {Van~Deventer}\ and\ \citenamefont
  {Wittmann}(2007)}]{van2007comparing}%
  \BibitemOpen
  \bibfield  {author} {\bibinfo {author} {\bibfnamefont {J.}~\bibnamefont
  {Van~Deventer}}\ and\ \bibinfo {author} {\bibfnamefont {M.~C.}\ \bibnamefont
  {Wittmann}},\ }\bibfield  {title} {\bibinfo {title} {Comparing student use of
  mathematical and physical vector representations},\ }in\ \href@noop {} {\emph
  {\bibinfo {booktitle} {AIP Conference Proceedings}}},\ Vol.\ \bibinfo
  {volume} {951}\ (\bibinfo {organization} {American Institute of Physics},\
  \bibinfo {year} {2007})\ pp.\ \bibinfo {pages} {208--211}\BibitemShut
  {NoStop}%
\bibitem [{\citenamefont {Aguirre}\ and\ \citenamefont
  {Erickson}(1984)}]{aguirre1984students}%
  \BibitemOpen
  \bibfield  {author} {\bibinfo {author} {\bibfnamefont {J.}~\bibnamefont
  {Aguirre}}\ and\ \bibinfo {author} {\bibfnamefont {G.}~\bibnamefont
  {Erickson}},\ }\bibfield  {title} {\bibinfo {title} {Students' conceptions
  about the vector characteristics of three physics concepts},\ }\href@noop {}
  {\bibfield  {journal} {\bibinfo  {journal} {Journal of research in Science
  Teaching}\ }\textbf {\bibinfo {volume} {21}},\ \bibinfo {pages} {439}
  (\bibinfo {year} {1984})}\BibitemShut {NoStop}%
\bibitem [{\citenamefont {Aguirre}(1988)}]{aguirre1988student}%
  \BibitemOpen
  \bibfield  {author} {\bibinfo {author} {\bibfnamefont {J.~M.}\ \bibnamefont
  {Aguirre}},\ }\bibfield  {title} {\bibinfo {title} {Student preconceptions
  about vector kinematics},\ }\href@noop {} {\bibfield  {journal} {\bibinfo
  {journal} {The Physics Teacher}\ }\textbf {\bibinfo {volume} {26}},\ \bibinfo
  {pages} {212} (\bibinfo {year} {1988})}\BibitemShut {NoStop}%
\bibitem [{\citenamefont {Barniol}\ \emph {et~al.}(2013)\citenamefont
  {Barniol}, \citenamefont {Zavala},\ and\ \citenamefont
  {Hinojosa}}]{barniol2013students}%
  \BibitemOpen
  \bibfield  {author} {\bibinfo {author} {\bibfnamefont {P.}~\bibnamefont
  {Barniol}}, \bibinfo {author} {\bibfnamefont {G.}~\bibnamefont {Zavala}},\
  and\ \bibinfo {author} {\bibfnamefont {C.}~\bibnamefont {Hinojosa}},\
  }\bibfield  {title} {\bibinfo {title} {Students' difficulties in interpreting
  the torque vector in a physical situation},\ }in\ \href@noop {} {\emph
  {\bibinfo {booktitle} {AIP Conference Proceedings}}},\ Vol.\ \bibinfo
  {volume} {1513}\ (\bibinfo {organization} {American Institute of Physics},\
  \bibinfo {year} {2013})\ pp.\ \bibinfo {pages} {58--61}\BibitemShut {NoStop}%
\bibitem [{\citenamefont {Liu}\ and\ \citenamefont
  {Mohattala}(2021)}]{liu2021study}%
  \BibitemOpen
  \bibfield  {author} {\bibinfo {author} {\bibfnamefont {D.}~\bibnamefont
  {Liu}}\ and\ \bibinfo {author} {\bibfnamefont {H.}~\bibnamefont
  {Mohattala}},\ }\bibfield  {title} {\bibinfo {title} {A study of
  undergraduates’ understanding of vector-decomposition of forces on inclined
  planes},\ }in\ \href@noop {} {\emph {\bibinfo {booktitle} {Proceedings of the
  Physics Education Research Conference (PERC}}}\ (\bibinfo {year} {2021})\
  pp.\ \bibinfo {pages} {233--238}\BibitemShut {NoStop}%
\bibitem [{\citenamefont {Rainson}\ \emph {et~al.}(1994)\citenamefont
  {Rainson}, \citenamefont {Transtr{\"o}mer},\ and\ \citenamefont
  {Viennot}}]{rainson1994students}%
  \BibitemOpen
  \bibfield  {author} {\bibinfo {author} {\bibfnamefont {S.}~\bibnamefont
  {Rainson}}, \bibinfo {author} {\bibfnamefont {G.}~\bibnamefont
  {Transtr{\"o}mer}},\ and\ \bibinfo {author} {\bibfnamefont {L.}~\bibnamefont
  {Viennot}},\ }\bibfield  {title} {\bibinfo {title} {Students’ understanding
  of superposition of electric fields},\ }\href@noop {} {\bibfield  {journal}
  {\bibinfo  {journal} {American Journal of Physics}\ }\textbf {\bibinfo
  {volume} {62}},\ \bibinfo {pages} {1026} (\bibinfo {year}
  {1994})}\BibitemShut {NoStop}%
\bibitem [{\citenamefont {Shaffer}\ and\ \citenamefont
  {McDermott}(2005)}]{shaffer2005research}%
  \BibitemOpen
  \bibfield  {author} {\bibinfo {author} {\bibfnamefont {P.~S.}\ \bibnamefont
  {Shaffer}}\ and\ \bibinfo {author} {\bibfnamefont {L.~C.}\ \bibnamefont
  {McDermott}},\ }\bibfield  {title} {\bibinfo {title} {A research-based
  approach to improving student understanding of the vector nature of
  kinematical concepts},\ }\href@noop {} {\bibfield  {journal} {\bibinfo
  {journal} {American Journal of Physics}\ }\textbf {\bibinfo {volume} {73}},\
  \bibinfo {pages} {921} (\bibinfo {year} {2005})}\BibitemShut {NoStop}%
\bibitem [{\citenamefont {Southey}\ and\ \citenamefont
  {Allie}(2014)}]{southey2014vector}%
  \BibitemOpen
  \bibfield  {author} {\bibinfo {author} {\bibfnamefont {P.}~\bibnamefont
  {Southey}}\ and\ \bibinfo {author} {\bibfnamefont {S.}~\bibnamefont
  {Allie}},\ }\bibfield  {title} {\bibinfo {title} {Vector addition in
  different contexts},\ }in\ \href@noop {} {\emph {\bibinfo {booktitle} {2014
  Physics Education Research Conference Proceedings}}}\ (\bibinfo {year}
  {2014})\ pp.\ \bibinfo {pages} {243--246}\BibitemShut {NoStop}%
\bibitem [{\citenamefont {Viennot}\ and\ \citenamefont
  {Rainson}(1992)}]{viennot1992students}%
  \BibitemOpen
  \bibfield  {author} {\bibinfo {author} {\bibfnamefont {L.}~\bibnamefont
  {Viennot}}\ and\ \bibinfo {author} {\bibfnamefont {S.}~\bibnamefont
  {Rainson}},\ }\bibfield  {title} {\bibinfo {title} {Students’ reasoning
  about the superposition of electric fields},\ }\href@noop {} {\bibfield
  {journal} {\bibinfo  {journal} {International Journal of Science Education}\
  }\textbf {\bibinfo {volume} {14}},\ \bibinfo {pages} {475} (\bibinfo {year}
  {1992})}\BibitemShut {NoStop}%
\bibitem [{\citenamefont {Smith}\ \emph {et~al.}(2018)\citenamefont {Smith},
  \citenamefont {McNeil}, \citenamefont {Guynn}, \citenamefont {Churukian},
  \citenamefont {Deardorff},\ and\ \citenamefont
  {Wallace}}]{smith2018transforming}%
  \BibitemOpen
  \bibfield  {author} {\bibinfo {author} {\bibfnamefont {D.~P.}\ \bibnamefont
  {Smith}}, \bibinfo {author} {\bibfnamefont {L.~E.}\ \bibnamefont {McNeil}},
  \bibinfo {author} {\bibfnamefont {D.~T.}\ \bibnamefont {Guynn}}, \bibinfo
  {author} {\bibfnamefont {A.~D.}\ \bibnamefont {Churukian}}, \bibinfo {author}
  {\bibfnamefont {D.~L.}\ \bibnamefont {Deardorff}},\ and\ \bibinfo {author}
  {\bibfnamefont {C.~S.}\ \bibnamefont {Wallace}},\ }\bibfield  {title}
  {\bibinfo {title} {Transforming the content, pedagogy and structure of an
  introductory physics course for life sciences majors},\ }\href@noop {}
  {\bibfield  {journal} {\bibinfo  {journal} {American Journal of Physics}\
  }\textbf {\bibinfo {volume} {86}},\ \bibinfo {pages} {862} (\bibinfo {year}
  {2018})}\BibitemShut {NoStop}%
\bibitem [{\citenamefont {Manogue}\ \emph {et~al.}(2006)\citenamefont
  {Manogue}, \citenamefont {Browne}, \citenamefont {Dray},\ and\ \citenamefont
  {Edwards}}]{manogue2006ampere}%
  \BibitemOpen
  \bibfield  {author} {\bibinfo {author} {\bibfnamefont {C.~A.}\ \bibnamefont
  {Manogue}}, \bibinfo {author} {\bibfnamefont {K.}~\bibnamefont {Browne}},
  \bibinfo {author} {\bibfnamefont {T.}~\bibnamefont {Dray}},\ and\ \bibinfo
  {author} {\bibfnamefont {B.}~\bibnamefont {Edwards}},\ }\bibfield  {title}
  {\bibinfo {title} {Why is amp{\`e}re’s law so hard? a look at
  middle-division physics},\ }\href@noop {} {\bibfield  {journal} {\bibinfo
  {journal} {American Journal of Physics}\ }\textbf {\bibinfo {volume} {74}},\
  \bibinfo {pages} {344} (\bibinfo {year} {2006})}\BibitemShut {NoStop}%
\bibitem [{\citenamefont {Reif}(1987)}]{reif1987interpretation}%
  \BibitemOpen
  \bibfield  {author} {\bibinfo {author} {\bibfnamefont {F.}~\bibnamefont
  {Reif}},\ }\bibfield  {title} {\bibinfo {title} {Interpretation of scientific
  or mathematical concepts: Cognitive issues and instructional implications},\
  }\href@noop {} {\bibfield  {journal} {\bibinfo  {journal} {Cognitive
  Science}\ }\textbf {\bibinfo {volume} {11}},\ \bibinfo {pages} {395}
  (\bibinfo {year} {1987})}\BibitemShut {NoStop}%
\bibitem [{\citenamefont {Otero}\ and\ \citenamefont
  {Harlow}(2009)}]{otero2009getting}%
  \BibitemOpen
  \bibfield  {author} {\bibinfo {author} {\bibfnamefont {V.~K.}\ \bibnamefont
  {Otero}}\ and\ \bibinfo {author} {\bibfnamefont {D.~B.}\ \bibnamefont
  {Harlow}},\ }\bibfield  {title} {\bibinfo {title} {Getting started in
  qualitative physics education research},\ }\href@noop {} {\bibfield
  {journal} {\bibinfo  {journal} {Reviews in PER}\ }\textbf {\bibinfo {volume}
  {2}} (\bibinfo {year} {2009})}\BibitemShut {NoStop}%
\bibitem [{\citenamefont {Willis}(2005)}]{willis2005cognitive}%
  \BibitemOpen
  \bibfield  {author} {\bibinfo {author} {\bibfnamefont {G.~B.}\ \bibnamefont
  {Willis}},\ }\bibfield  {title} {\bibinfo {title} {Cognitive interviewing in
  practice: think-aloud, verbal probing and other techniques},\ }\href@noop {}
  {\bibfield  {journal} {\bibinfo  {journal} {Cognitive interviewing: a tool
  for improving questionnaire design. London: Sage Publications}\ ,\ \bibinfo
  {pages} {42}} (\bibinfo {year} {2005})}\BibitemShut {NoStop}%
\bibitem [{\citenamefont {Barbieri}\ and\ \citenamefont
  {Booth}(2020)}]{barbieri2020mistakes}%
  \BibitemOpen
  \bibfield  {author} {\bibinfo {author} {\bibfnamefont {C.~A.}\ \bibnamefont
  {Barbieri}}\ and\ \bibinfo {author} {\bibfnamefont {J.~L.}\ \bibnamefont
  {Booth}},\ }\bibfield  {title} {\bibinfo {title} {Mistakes on display:
  Incorrect examples refine equation solving and algebraic feature knowledge},\
  }\href@noop {} {\bibfield  {journal} {\bibinfo  {journal} {Applied Cognitive
  Psychology}\ }\textbf {\bibinfo {volume} {34}},\ \bibinfo {pages} {862}
  (\bibinfo {year} {2020})}\BibitemShut {NoStop}%
\bibitem [{\citenamefont {De~La~Hoz}\ \emph {et~al.}(2023)\citenamefont
  {De~La~Hoz}, \citenamefont {Vieira},\ and\ \citenamefont
  {Arteta}}]{de2023self}%
  \BibitemOpen
  \bibfield  {author} {\bibinfo {author} {\bibfnamefont {J.~L.}\ \bibnamefont
  {De~La~Hoz}}, \bibinfo {author} {\bibfnamefont {C.}~\bibnamefont {Vieira}},\
  and\ \bibinfo {author} {\bibfnamefont {C.}~\bibnamefont {Arteta}},\
  }\bibfield  {title} {\bibinfo {title} {Self-explanation activities in
  statics: A knowledge-building activity to promote conceptual change},\
  }\href@noop {} {\bibfield  {journal} {\bibinfo  {journal} {Journal of
  Engineering Education}\ }\textbf {\bibinfo {volume} {112}},\ \bibinfo {pages}
  {741} (\bibinfo {year} {2023})}\BibitemShut {NoStop}%
\bibitem [{\citenamefont {Maclean}\ and\ \citenamefont
  {Bayley}(2023)}]{maclean2023s}%
  \BibitemOpen
  \bibfield  {author} {\bibinfo {author} {\bibfnamefont {K.~D.}\ \bibnamefont
  {Maclean}}\ and\ \bibinfo {author} {\bibfnamefont {T.}~\bibnamefont
  {Bayley}},\ }\bibfield  {title} {\bibinfo {title} {That’s incorrect and let
  me tell you why: A scalable assessment to evaluate higher order thinking
  skills},\ }\href@noop {} {\bibfield  {journal} {\bibinfo  {journal} {INFORMS
  Transactions on Education}\ } (\bibinfo {year} {2023})}\BibitemShut {NoStop}%
\bibitem [{\citenamefont {Yerushalmi}\ and\ \citenamefont
  {Polingher}(2006)}]{yerushalmi2006guiding}%
  \BibitemOpen
  \bibfield  {author} {\bibinfo {author} {\bibfnamefont {E.}~\bibnamefont
  {Yerushalmi}}\ and\ \bibinfo {author} {\bibfnamefont {C.}~\bibnamefont
  {Polingher}},\ }\bibfield  {title} {\bibinfo {title} {Guiding students to
  learn from mistakes},\ }\href@noop {} {\bibfield  {journal} {\bibinfo
  {journal} {Physics Education}\ }\textbf {\bibinfo {volume} {41}},\ \bibinfo
  {pages} {532} (\bibinfo {year} {2006})}\BibitemShut {NoStop}%
\bibitem [{\citenamefont {McDermott}\ \emph {et~al.}(2002)\citenamefont
  {McDermott}, \citenamefont {Shaffer} \emph
  {et~al.}}]{mcdermott2002tutorials}%
  \BibitemOpen
  \bibfield  {author} {\bibinfo {author} {\bibfnamefont {L.~C.}\ \bibnamefont
  {McDermott}}, \bibinfo {author} {\bibfnamefont {P.~S.}\ \bibnamefont
  {Shaffer}}, \emph {et~al.},\ }\href@noop {} {\emph {\bibinfo {title}
  {Tutorials in Introductory Physics}}}\ (\bibinfo  {publisher} {Prentice Hall
  Upper Saddle River, NJ},\ \bibinfo {year} {2002})\BibitemShut {NoStop}%
\bibitem [{\citenamefont {Prather}\ \emph {et~al.}(2022)\citenamefont
  {Prather}, \citenamefont {Brissenden}, \citenamefont {Wallace},\ and\
  \citenamefont {Adams}}]{prather2022lecture}%
  \BibitemOpen
  \bibfield  {author} {\bibinfo {author} {\bibfnamefont {E.~E.}\ \bibnamefont
  {Prather}}, \bibinfo {author} {\bibfnamefont {G.}~\bibnamefont {Brissenden}},
  \bibinfo {author} {\bibfnamefont {C.~S.}\ \bibnamefont {Wallace}},\ and\
  \bibinfo {author} {\bibfnamefont {J.~P.}\ \bibnamefont {Adams}},\ }\href@noop
  {} {\emph {\bibinfo {title} {Lecture-Tutorials for Introductory
  Astronomy}}},\ \bibinfo {edition} {4th}\ ed.\ (\bibinfo  {publisher} {Pearson
  Education, Inc, San Francisco, CA},\ \bibinfo {year} {2022})\BibitemShut
  {NoStop}%
\bibitem [{\citenamefont {Bloom}\ and\ \citenamefont
  {Krathwohl}(2020)}]{bloom2020taxonomy}%
  \BibitemOpen
  \bibfield  {author} {\bibinfo {author} {\bibfnamefont {B.~S.}\ \bibnamefont
  {Bloom}}\ and\ \bibinfo {author} {\bibfnamefont {D.~R.}\ \bibnamefont
  {Krathwohl}},\ }\href@noop {} {\emph {\bibinfo {title} {Taxonomy of
  educational objectives: The classification of educational goals. Book 1,
  Cognitive domain}}}\ (\bibinfo  {publisher} {Longman},\ \bibinfo {year}
  {2020})\BibitemShut {NoStop}%
\bibitem [{\citenamefont {Wallace}\ \emph {et~al.}(2021)\citenamefont
  {Wallace}, \citenamefont {Prather}, \citenamefont {Milsom}, \citenamefont
  {Johns},\ and\ \citenamefont {Manne}}]{wallace2021students}%
  \BibitemOpen
  \bibfield  {author} {\bibinfo {author} {\bibfnamefont {C.~S.}\ \bibnamefont
  {Wallace}}, \bibinfo {author} {\bibfnamefont {E.~E.}\ \bibnamefont
  {Prather}}, \bibinfo {author} {\bibfnamefont {J.~A.}\ \bibnamefont {Milsom}},
  \bibinfo {author} {\bibfnamefont {K.}~\bibnamefont {Johns}},\ and\ \bibinfo
  {author} {\bibfnamefont {S.}~\bibnamefont {Manne}},\ }\bibfield  {title}
  {\bibinfo {title} {Students taught by a first-time instructor using
  active-learning teaching strategies outperform students taught by a
  highly-regarded traditional instructor},\ }\href@noop {} {\bibfield
  {journal} {\bibinfo  {journal} {Journal of College Science Teaching}\
  }\textbf {\bibinfo {volume} {50}},\ \bibinfo {pages} {48} (\bibinfo {year}
  {2021})}\BibitemShut {NoStop}%
\bibitem [{\citenamefont {Wallace}(2020)}]{wallace2020developing}%
  \BibitemOpen
  \bibfield  {author} {\bibinfo {author} {\bibfnamefont {C.~S.}\ \bibnamefont
  {Wallace}},\ }\bibfield  {title} {\bibinfo {title} {Developing peer
  instruction questions for quantitative problems for an upper-division
  astronomy course},\ }\href@noop {} {\bibfield  {journal} {\bibinfo  {journal}
  {American Journal of Physics}\ }\textbf {\bibinfo {volume} {88}},\ \bibinfo
  {pages} {214} (\bibinfo {year} {2020})}\BibitemShut {NoStop}%
\bibitem [{\citenamefont {Schoenfeld}(1987)}]{schoenfeld1987s}%
  \BibitemOpen
  \bibfield  {author} {\bibinfo {author} {\bibfnamefont {A.~H.}\ \bibnamefont
  {Schoenfeld}},\ }\bibfield  {title} {\bibinfo {title} {What's all the fuss
  about metacognition?},\ }\href@noop {} {\bibfield  {journal} {\bibinfo
  {journal} {Cognitive science and mathematics education}\ ,\ \bibinfo {pages}
  {189}} (\bibinfo {year} {1987})}\BibitemShut {NoStop}%
\bibitem [{\citenamefont {Hsu}\ \emph {et~al.}(2004)\citenamefont {Hsu},
  \citenamefont {Brewe}, \citenamefont {Foster},\ and\ \citenamefont
  {Harper}}]{hsu2004resource}%
  \BibitemOpen
  \bibfield  {author} {\bibinfo {author} {\bibfnamefont {L.}~\bibnamefont
  {Hsu}}, \bibinfo {author} {\bibfnamefont {E.}~\bibnamefont {Brewe}}, \bibinfo
  {author} {\bibfnamefont {T.~M.}\ \bibnamefont {Foster}},\ and\ \bibinfo
  {author} {\bibfnamefont {K.~A.}\ \bibnamefont {Harper}},\ }\bibfield  {title}
  {\bibinfo {title} {Resource letter rps-1: Research in problem solving},\
  }\href@noop {} {\bibfield  {journal} {\bibinfo  {journal} {American Journal
  of Physics}\ }\textbf {\bibinfo {volume} {72}},\ \bibinfo {pages} {1147}
  (\bibinfo {year} {2004})}\BibitemShut {NoStop}%
\bibitem [{\citenamefont {Leonard}\ \emph {et~al.}(1996)\citenamefont
  {Leonard}, \citenamefont {Dufresne},\ and\ \citenamefont
  {Mestre}}]{leonard1996using}%
  \BibitemOpen
  \bibfield  {author} {\bibinfo {author} {\bibfnamefont {W.~J.}\ \bibnamefont
  {Leonard}}, \bibinfo {author} {\bibfnamefont {R.~J.}\ \bibnamefont
  {Dufresne}},\ and\ \bibinfo {author} {\bibfnamefont {J.~P.}\ \bibnamefont
  {Mestre}},\ }\bibfield  {title} {\bibinfo {title} {Using qualitative
  problem-solving strategies to highlight the role of conceptual knowledge in
  solving problems},\ }\href@noop {} {\bibfield  {journal} {\bibinfo  {journal}
  {American Journal of Physics}\ }\textbf {\bibinfo {volume} {64}},\ \bibinfo
  {pages} {1495} (\bibinfo {year} {1996})}\BibitemShut {NoStop}%
\bibitem [{\citenamefont {Sabella}\ and\ \citenamefont
  {Redish}(2007)}]{sabella2007knowledge}%
  \BibitemOpen
  \bibfield  {author} {\bibinfo {author} {\bibfnamefont {M.~S.}\ \bibnamefont
  {Sabella}}\ and\ \bibinfo {author} {\bibfnamefont {E.~F.}\ \bibnamefont
  {Redish}},\ }\bibfield  {title} {\bibinfo {title} {Knowledge organization and
  activation in physics problem solving},\ }\href@noop {} {\bibfield  {journal}
  {\bibinfo  {journal} {American Journal of Physics}\ }\textbf {\bibinfo
  {volume} {75}},\ \bibinfo {pages} {1017} (\bibinfo {year}
  {2007})}\BibitemShut {NoStop}%
\bibitem [{\citenamefont {Wilcox}\ \emph {et~al.}(2013)\citenamefont {Wilcox},
  \citenamefont {Caballero}, \citenamefont {Rehn},\ and\ \citenamefont
  {Pollock}}]{wilcox2013analytic}%
  \BibitemOpen
  \bibfield  {author} {\bibinfo {author} {\bibfnamefont {B.~R.}\ \bibnamefont
  {Wilcox}}, \bibinfo {author} {\bibfnamefont {M.~D.}\ \bibnamefont
  {Caballero}}, \bibinfo {author} {\bibfnamefont {D.~A.}\ \bibnamefont
  {Rehn}},\ and\ \bibinfo {author} {\bibfnamefont {S.~J.}\ \bibnamefont
  {Pollock}},\ }\bibfield  {title} {\bibinfo {title} {Analytic framework for
  students’ use of mathematics in upper-division physics},\ }\href@noop {}
  {\bibfield  {journal} {\bibinfo  {journal} {Physical Review Special
  Topics-Physics Education Research}\ }\textbf {\bibinfo {volume} {9}},\
  \bibinfo {pages} {020119} (\bibinfo {year} {2013})}\BibitemShut {NoStop}%
\end{thebibliography}%
\nocite{*}
\end{document}